\documentclass[12pt]{iopart}

\usepackage{graphicx}
\usepackage{layout}
\begin{document}

\title[Lorentz contraction and accelerated systems]{Lorentz contraction and accelerated systems}

\author{Angelo Tartaglia$^{\ast,\S}$\ and Matteo Luca
Ruggiero$^{\ast,\S}$}

\address{email: angelo.tartaglia@polito.it, \ matteo.ruggiero@polito.it}
\address{$^\ast$\ Dipartimento di Fisica, Politecnico di Torino,
 Corso Duca degli Abruzzi 24, 10129 Torino, Italy }

\address{$^\S$\ INFN, Via P. Giuria 1, 10125 Torino, Italy}

\begin{abstract}
The paper discusses the problem of the Lorentz contraction in
accelerated systems, in the context of the special theory of
relativity. Equal proper accelerations along different world lines
are considered, showing the differences arising when the world
lines correspond to physically connected or disconnected objects.
In all cases the special theory of relativity proves to be
completely self-consistent.\end{abstract}


\maketitle

\section{Introduction}

The behavior of accelerated systems in special relativity is a
delicate problem, that deserves a careful analysis lest apparent
paradoxes and contradictions seem to come up. Usually more
attention is paid in textbooks to kinematical 'paradoxes', whilst
accelerated systems are not discussed at length. This at least is
what emerges from reading such classic texts as \cite{rindler1},
\cite{rindler}, \cite{landau}, \cite{ruffi}, \cite{bohm},
\cite{rosser}, \cite{taylor}, \cite{katz}, \cite{adler}.

Of course acceleration plus the equivalence principle is a
fundamental and delicate issue, considering that gravitation and
acceleration are locally indistinguishable. We think that
accelerated systems should be fully discussed in order to obtain a
good understanding of special relativity.

In this paper a simple problem will be considered which, at a
superficial glance, can disorient students. The final conclusion
will be that special relativity, once again, is self consistent.

\section{Posing the problem}

Two objects at different positions along the $x$ axis of an
inertial reference frame undergo equal accelerations during equal
coordinate time intervals. In this condition we expect that the
distance between them, seen in the initial rest frame, always
remains equal to its initial (rest) value. At the end of the
accelerated phase however, when both bodies will move with the
same constant speed, the distance between them should turn out to
be Lorentz contracted with respect to the rest (proper) value. Is
there a contradiction? Has the proper distance increased during
the acceleration time? If the two objects are physically
connected, has a stress set in?

\subsection{One observer, one meter}

Suppose, to begin with, that there is a fixed inertial observer at
the origin; let us call him $O$. Then add a second observer
$O^{\prime }$, initially coincident with $O$; both observers carry
equal meter rods, stretched along the $x$ axis; the length of the
rods in the rest inertial frame is $l$. $O^{\prime }$ is set into
accelerated motion along the $x$ axis starting at $t=0$; let us
assume, for the sake of simplicity, that the proper acceleration
of $O^{\prime }$ is a constant $a$.

The world line of $O^{\prime }$ in the reference frame of $O$ is a
hyperbola described by the equation \cite{misner}
\begin{equation}
\left( x+\frac{c^{2}}{a}\right)
^{2}-c^{2}t^{2}=\frac{c^{4}}{a^{2}} \label{hiperbola}
\end{equation}

The metric in the frame of $O$ is of course (letting aside the
irrelevant $y$ and $z$ dimensions):
\[
ds^{2}=c^{2}dt^{2}-dx^{2}
\]%
From the view point of $O^{\prime }$ the acceleration produces the
same
effect as a uniform gravitational field along $x$\footnote{%
The issue of the local equivalence between accelerated systems and
gravitational fields is not a trivial one (see for instance
\cite{desloge} and \cite{wheeler}). In our case however the
situation is simple if we assume that a positive acceleration is
applied to the rear end of the rod, which is otherwise free. The
possibility of using an infinite set of inertial frames to
describe an accelerated observer deserves in turn a careful
discussion.In our case however all these problems are not relevant
for the final conclusions.}, consequently a rod aligned with $x$
will be compressed and accordingly shortened with respect to a rod
along, say, $y$. The rigid rod case has been discussed in
\cite{rindler66}. In our case no rigidity is assumed for the
extended rods since it would contradict the relativity theory. The
amount of the deformation of the rod along $x$ will depend on the
nature of the rod and, specifically, on the stiffness of the
material which it is made of. Calling $k$ the stiffness of the rod and $%
\lambda $ the proper density (per unit length) of its material, then $%
l^{\prime }$, which is the length of the rod as seen by $O^{\prime
}$, turns
out to be\footnote{%
The rod, in the uniformly accelerated frame, is in equilibrium
under the action of the elastic force and the gravitational-like
force along $x$, due to its acceleration. Hence, the total
potential energy can be written as
\[
W=\frac{1}{2}k(l^{\prime }-l)^{2}+\frac{1}{2}\lambda al^{\prime
}{}^{2}
\]%
where $\lambda $ is assumed to be a constant. Differentiating with
respect to $l^{\prime }$, formula (\ref{elasti}) is obtained.}
\begin{equation}
l^{\prime }=l\frac{k}{k+\lambda a}  \label{elasti}
\end{equation}%
The length seen by $O$ is obtained from $l^{\prime }$ projecting
it from the $x^{\prime }$ axis (the space of $O^{\prime }$) to the
$x$ axis, i.e. multiplying by $\sqrt{1-v^{2}/c^{2}}$, according to
the standard Lorentz contraction. The instantaneous coordinate
velocity $v=\frac{dx}{dt}$, as seen by $O$, is calculated from
(\ref{hiperbola}):
\begin{equation}
v=\frac{ct}{\sqrt{\left( \frac{c^{2}}{a^{2}}+t^{2}\right) }}=\frac{at}{\sqrt{%
1+\frac{a^{2}t^{2}}{c^{2}}}}  \label{velocity}
\end{equation}

Finally the length seen by $O$ will be
\[
l^{\prime \prime }=l^{\prime }\sqrt{1-\frac{v^{2}}{c^{2}}}=\frac{l^{\prime }%
}{\sqrt{\left( 1+\frac{t^{2}a^{2}}{c^{2}}\right) }}\allowbreak =\frac{k}{%
k+\lambda a}\frac{l}{\sqrt{\left(
1+\frac{t^{2}a^{2}}{c^{2}}\right) }}
\]

The rod, in the frame of $O$, gets contracted more and more as
time passes. Considering this fact, since the effect depends on
the properties of material bodies, all lengths measured along the
$x$ direction using the rod appear distorted with respect to the
initial rest values. Nothing happens to the distances measured in
any transverse direction.

When the acceleration phase comes to an end at coordinate time
$t_{0}$, the translational velocity, with respect to $O$, keeps
its constant final value
\[
v_{0}=\frac{ct_{0}}{\sqrt{\left(
\frac{c^{2}}{a^{2}}+t_{0}^{2}\right) }}
\]
and $a$ drops to zero in (\ref{elasti}). Not considering
oscillations and internal energy dissipation, the rod recovers its
initial proper length, and the corresponding length seen by $O$ is
\[
l_{0}^{\prime \prime }=l\sqrt{1-v_{0}^{2}/c^{2}}
\]
The result is exactly what was expected to be: no paradox of any
sort appears, all measured lengths recover the initial
unaccelerated values.

\section{Light rays}

\begin{figure}[t]
\begin{center}
\includegraphics[width=10cm,height=8cm]{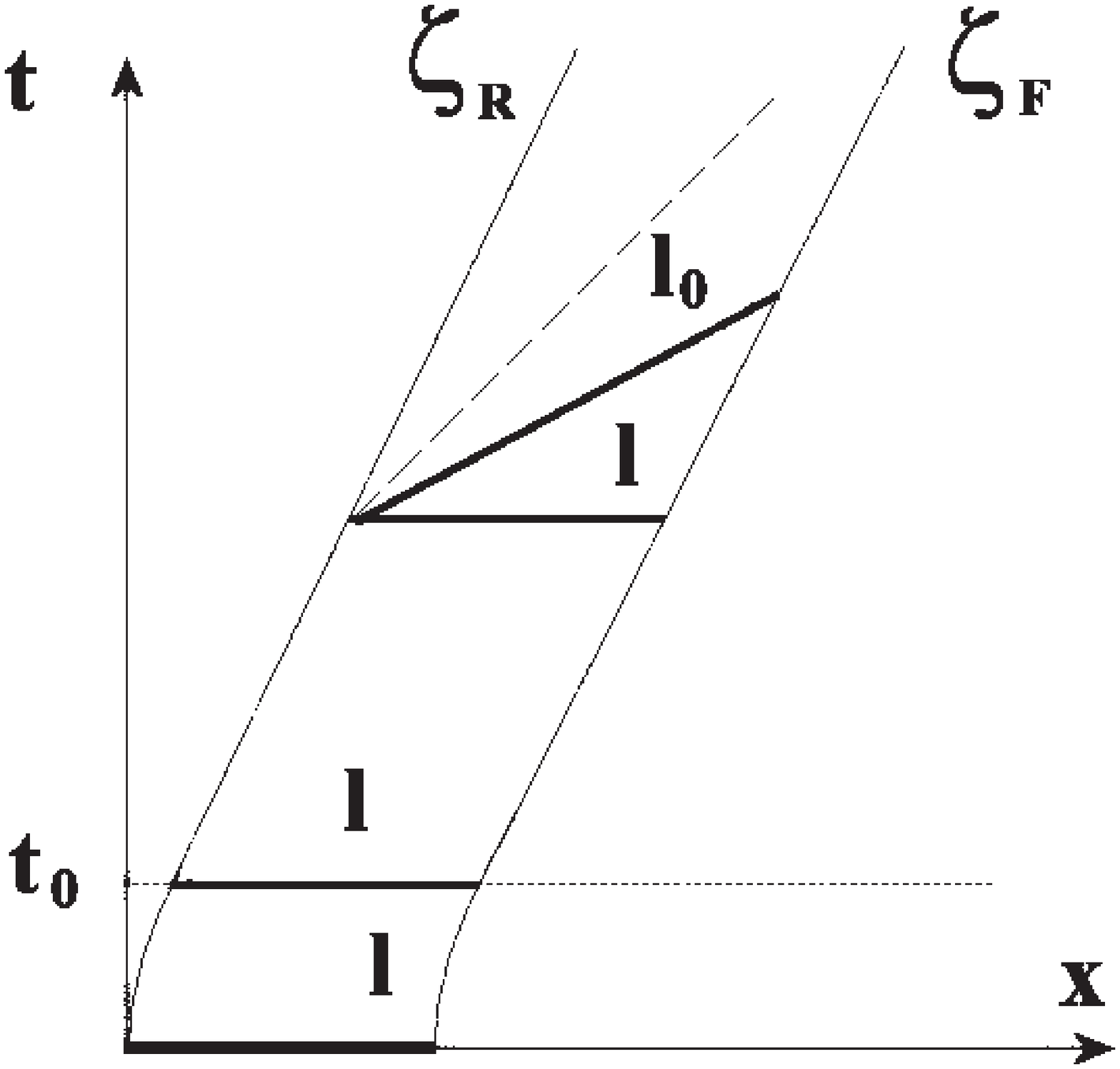}%
\end{center}
\caption{The world lines $\protect\zeta _{R}$ and $\protect\zeta
_{F}$ of two objects accelerated with the same proper acceleration
and for the same proper times up to $t=t_{0}$ are shown.
Afterwards both objects continue to move inertially at the same
constant speed. The distance between the objects as seen by a
static observer maintains its initial value $l$, but the proper
distance increases to become $l_{0}$. The length $l$ corresponds
to the Lorentz contracted $l_{0}$. The dashed line represents the
light cone.} \label{fig:figura1}
\end{figure}

Let us now consider a situation where two equal rockets are
initially placed on the $x$ axis at a distance $l$ from one
another. Every rocket carries on board a scientist to make
measurements, and an engineer to control the thrust of the rocket.
The engineers carry identical (initially) synchronized clocks and
have the same instructions for the regime of the engines. Let us
call $F$ the front rocket (and moving observer), and $R$ the rear
rocket, with its observer(see figure \ref{fig:figura1}). $F$ and
$R$ are not physically connected, so that they move exactly with
the same proper acceleration at any time. The way used to monitor
the reciprocal positions is the exchange of light rays.

The infinitesimal proper time interval $d\tau $ is given, in terms
of the coordinate time interval $dt$, by
\begin{equation}
d\tau =dt\sqrt{1-\frac{v^{2}}{c^{2}}}\equiv \gamma ^{-1}dt
\label{eq:proptime1}
\end{equation}
where we have introduced the Lorentz factor $\gamma (v)=1/\sqrt{1-\frac{v^{2}%
}{c^{2}}}$.\newline Hence, substituting $v$ from \ref{velocity},
we obtain
\begin{equation}
d\tau =\frac{dt}{\sqrt{1+\frac{a^{2}t^{2}}{c^{2}}}}
\label{eq:proptime2}
\end{equation}
in terms of the coordinate time. By integrating ($\tau =0$ when
$t=0$), the proper time lapse turns out to be
\begin{equation}
\tau =\frac{c}{a}\sinh ^{-1}\left( \frac{a}{c}t\right)
\label{eq:proptime3}
\end{equation}
hence, we obtain
\begin{equation}
t=\frac{c}{a}\sinh \left( \frac{a}{c}\tau \right)  \label{eq:time}
\end{equation}
Now, if we substitute in \ref{velocity} this expression of
coordinate time, as a function of the elapsed proper time, we see
that the coordinate velocity can be written as
\begin{equation}
v=c\tanh \left( \frac{a}{c}\tau \right)  \label{eq:vtau}
\end{equation}
At a given predetermined $\tau _{0}$ the engines are stopped on
both rockets. From that moment on, both for $R$ and $F$, the
flight continues at a constant coordinate speed
\begin{equation}
v_{0}=\frac{dx}{dt}=c\tanh \left( \frac{a}{c}\tau _{0}\right)
\label{vnot}
\end{equation}
and the corresponding Lorentz factor $\gamma (v_{0})$ is
\begin{equation}
\gamma (v_{0})=\cosh \left( \frac{a}{c}\tau _{0}\right)
\label{eq:gammavzero}
\end{equation}

The round trip of light (\cite{rindler}) between the space ships
corresponds to a coordinate time interval
\[
c\delta t=2l\cosh ^{2}\left( \frac{a}{c}\tau _{0}\right)
\]
and in terms of proper time of $R$ ($\delta t=\gamma (v_{0})\delta
\tau $)
\begin{equation}
c\delta \tau =2l\cosh \left( \frac{a}{c}\tau _{0}\right)
\label{proplength}
\end{equation}
The proper distance is usually defined as $l_{0}=\frac{c\delta
\tau }{2}$. On the other hand, the distance seen by a static
observer in these circumstances remains always equal to $l$, but
the proper distance in the frame of the rockets, $l_{0}$, has
progressively increased during the acceleration so that its
Lorentz contraction ($l=\gamma ^{-1}l_{0}$) produces precisely the
$l$ result. In fact, from (\ref{proplength}) one has
\begin{equation}
l_{0}\equiv \frac{c\delta \tau }{2}=l\cosh \left( \frac{a}{c}\tau
_{0}\right) \label{eq:dilatazione}
\end{equation}
which, considering (\ref{vnot}) or (\ref{eq:gammavzero}),
corresponds to
\begin{equation}
l_{0}=\frac{l}{\sqrt{1-\frac{v_{0}^{2}}{c^{2}}}}
\label{eq:dilatazione2}
\end{equation}

\section{Two physically connected accelerated observers}

\begin{figure}[t]
\begin{center}
\includegraphics[width=8cm,height=8cm]{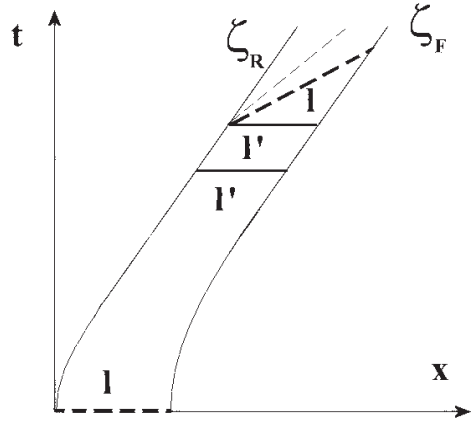}%
\end{center}
\caption{$\protect\zeta _{R}$ and $\protect\zeta _{F}$ represent
the world lines of two rockets connected by a spring, as explained
in the text. Once the acceleration finishes (at different
coordinate times for the two rockets) the ends of the spring will
keep on moving at a constant coordinate speed, and the spring will
recover its initial unstretched length $l$. A static observer will
measure the Lorentz contracted length $l^{\prime }$. }
\label{fig:figura2}
\end{figure}

Let us now consider a situation where both observers at $R$ and
$F$ are physically connected by a spring.

In this condition, one could expect the two ends of the rod to be
equally accelerated, however, just as in a gravitational field,
when trying to keep the length of a spring fixed notwithstanding
gravity, the spring will react with a pull on both ends, since its
rest length is now shorter than what it would be without
acceleration. The consequence will be that the actual acceleration
of the front end will be a little bit less than what the engine
alone would produce, and the acceleration of the rear end will be
a little bit more for the same reason. In this way, the proper
times of the two engineers will no longer be the same at a given
coordinate time and the two world lines of the ends of the spring
will no longer be equal hyperbolae (see figure \ref{fig:figura2}).
In fact, the rear world line will in general be more curved than
the front one. When the engines stop thrusting, at the same proper
times of the engineers, but at different coordinate times, after
some transient (including oscillations and dissipation of energy)
the situation will be such that the spring will remain unstretched
in its rest frame, i.e. its proper length will again be $l$ and of
course its both ends will move at the same coordinate speed. The
Earth bound observer will measure a properly contracted length
\[
l^{\prime }=l\sqrt{1-\frac{v^{2}}{c^{2}}}
\]

\section{Conclusion}

As we have seen by elementary considerations, the simple scheme of
two equally accelerated observers leads to results consistent with
the Lorentz contraction in any case. When the two observers are
physically connected, the material bridge between them insures the
classical contraction. When the two observers are not connected,
the relativity of simultaneity produces, from the view point of
the rear observer, a forward flight of the front observer, which
in the end will be seen as an increase in the proper distance,
leading, via Lorentz contraction, to a distance, in the frame of
the static observer, exactly equal to the initial one.

There is no need to analyze in detail the accelerated phase, when
such concepts as an extended proper distance are ill defined.
Actually the same conclusions are reached no matters what the
acceleration programs are, provided they are equal with respect to
proper times. The reason why we have considered constant proper
acceleration has been just for the sake of simplicity in
intermediate steps. We have also implicitly assumed that the sizes
of physical systems we considered, were small enough not to incur
into troubles with horizons and other difficulties typical of
extended accelerated reference frames \cite{misner}\cite{philpott}

We think that proposing an example/exercise of this sort to the
students would produce a deeper insight in the principles of
special relativity.

\end{document}